\documentclass[3p,times,twocolumn]{elsarticle}

\usepackage{ecrc}

\volume{00}

\firstpage{1}

\journalname{Nuclear Physics B Proceedings Supplement}

\runauth{J. Bortfeldt et al.}

\jid{nuphbp}

\jnltitlelogo{Nuclear Physics B Proceedings Supplement}

\usepackage{amssymb}

\usepackage[figuresright]{rotating}

\begin{document}

\begin{frontmatter}

\dochead{}

\title{High-Rate Capable Floating Strip Micromegas}

\author[label1]{Jonathan Bortfeldt}
\author[label1]{Michael Bender}
\author[label1]{Otmar Biebel}
\author[label1]{Helge Danger}
\author[label1]{Bernhard Flierl}
\author[label1]{Ralf Hertenberger}
\author[label1]{Philipp L\"osel}
\author[label1]{Samuel Moll}
\author[label2,label3]{Katia Parodi}
\author[label2,label3]{Ilaria Rinaldi}
\author[label1]{Alexander Ruschke}
\author[label4]{Andr\'e Zibell}
\address[label1]{LS Schaile, LMU Munich}
\address[label2]{LS Parodi, LMU Munich}
\address[label3]{Heidelberg University Hospital}
\address[label4]{University W\"urzburg}

\begin{abstract}
We report on the optimization of discharge insensitive floating strip Micromegas (MICRO-MEsh GASeous) detectors, fit for use in high-energy muon spectrometers. The suitability of these detectors for particle tracking is shown in high-background environments and at very high particle fluxes up to 60\,MHz/cm$^2$. Measurement and simulation of the microscopic discharge behavior have demonstrated the excellent discharge tolerance. A floating strip Micromegas with an active area of 48\,cm\,$\times$\,50\,cm with 1920 copper anode strips exhibits in 120\,GeV pion beams a spatial resolution of 50\,$\mu$m at detection efficiencies above 95\%. Pulse height, spatial resolution and detection efficiency are homogeneous over the detector. Reconstruction of particle track inclination in a single detector plane is discussed, optimum angular resolutions below $5^\circ$ are observed. Systematic deviations of this $\mu$TPC-method are fully understood. The reconstruction capabilities for minimum ionizing muons are investigated in a 6.4\,cm\,$\times$\,6.4\,cm floating strip Micromegas under intense background irradiation of the whole active area with 20\,MeV protons at a rate of 550\,kHz. The spatial resolution for muons is not distorted by space charge effects.
A 6.4\,cm\,$\times$\,6.4\,cm floating strip Micromegas doublet with low material budget is investigated in highly ionizing proton and carbon ion beams at particle rates between 2\,MHz and 2\,GHz. Stable operation up to the highest rates is observed, spatial resolution, detection efficiencies, the multi-hit and high-rate capability are discussed.
\end{abstract}

\begin{keyword}
tracking detectors \sep Micromegas \sep discharge protection \sep $\mu$TPC reconstruction \sep highest particle rates \sep high-rate background \sep ion transmission imaging

\end{keyword}

\end{frontmatter}


\section{Introduction}

Micromegas are planar micro-structured gaseous particle detectors. By exploiting short ion drift times, enabled by gas amplification in a thin region between a fine micro-mesh and anode strips, they are high-rate capable. Introduced in 1996 by Giomataris and Charpak \cite{giomataris:mm}, extensive research and development has led to the application of Micromegas detectors in several experiments in the past years \citep{thers:compassmm}, \citep{abgrall:t2ktpc}, \citep{dafni:castnewmm}.
Due to the elevated background hit rates in the Small Wheel Region of the ATLAS muon spectrometer after the LHC luminosity upgrade to $5\times10^{34}$\,cm$^{-2}$s$^{-1}$, Micromegas have been chosen as replacement for the currently used Monitored Drift Tube and Cathode Strip precision tracking chambers \cite{atlas:nswtdr}. 

Discharges between micro-mesh and anode strips can occur in Micromegas detectors. Although non-destructive, they create dead time due to the necessary restoration of the amplification field between mesh and anode strips. The efficiency as well as the spatial resolution can be degraded. Measurements have shown, that a local charge density in a $100$\,$\mu$m $\times$ $100$\,$\mu$m region exceeding $1.8\times10^{6}$\,e leads to the formation of streamers and subsequent discharges \cite{moll:da}. Minimum ionizing particles usually create ionization charges on the order of 100\,e, but the critical charge density can be exceeded by strongly ionizing particles such as alpha particles, low-energy protons or products from hadronic or electromagnetic showers. 

A powerful discharge suppression mechanism is realized in resistive strip Micromegas \cite{wotschack:mmresist}. Resistive anode strips with a resistivity of about 10\,M$\Omega$/cm cover the copper readout strips. In a discharge the potential of the affected resistive strip is changed only locally, such that the discharge ends very quickly as the electric field breaks down locally due to equilibration of the potential between strip and mesh.
\begin{figure}
\centering
\includegraphics[width=0.79\linewidth]{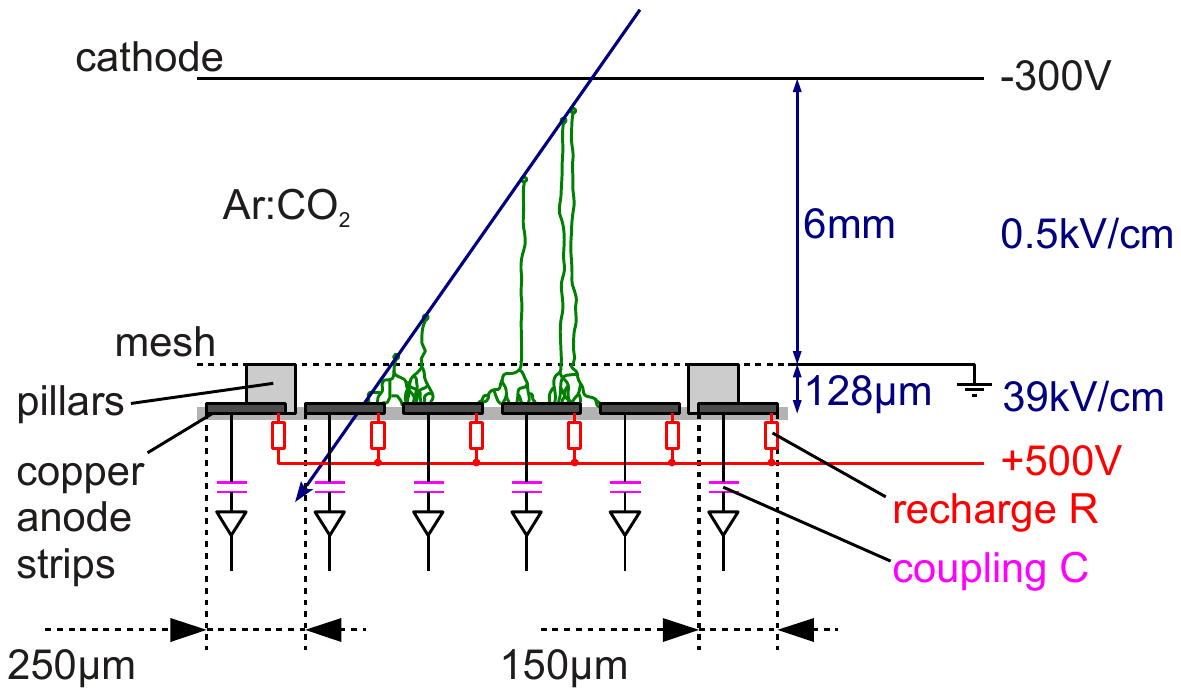}
\caption{Functional principle of Floating Strip Micromegas.}
\label{fig:MicomPrinzip}
\end{figure}

In this paper we discuss the development and the performance of an alternative discharge suppression configuration, called floating strip Micromegas. An early stage of the concept, has been proposed by \cite{thers:compassmm}, \cite{bay:mmsparking} and \cite{kane:newroscheme}. The amplification region is formed by a micro-mesh, held at ground potential, and copper anode strips that are individually connected to positive high-voltage of around 500\,V via $R \gtrsim 20$\,M$\Omega$ resistors (Figure \ref{fig:MicomPrinzip}). Small capacitors decouple the strips from the readout electronics without blocking signals. Since their capacitance is of the same order as the small strip-to-mesh, strip-to-strip and strip-to-ground capacitance, strips can undergo fast variations in their potential. Although the small coupling capacitance leads to a loss of signal, a powerful discharge suppression is enabled, compared to the standard, non-resistive Micromegas configuration, where the whole detector is affected during a discharge. First, the potential between the mesh and the few affected strips levels quickly in a discharge and disrupts the conductive connection. Only up to three strips are affected, the remaining strips are fully efficient and the global voltage drop on the common strip potential is negligibly small. Second, the recharge time is small, due to the small capacitance, regardless of the large recharge resistor. The capacitance and resistance values have been optimized in detailed simulation and experimental studies \cite{bortfeldt:phd}.

In the following we will show, that the floating strip mechanism works as expected. We will discuss the performance of floating strip Micromegas in three different measurement campaigns and thus demonstrate the suitability as particle tracking detector with excellent spatial resolution in large-area and high-rate application in high energy and medical physics. In all three campaigns, the Micromegas were flushed with an Ar:CO$_2$ 93:7 vol.~\% gas mixture at fluxes of 2.0\,ln/h.

\section{Setup and Assembly of the Detectors}
Three different types of floating strip Micromegas were constructed and investigated. 

\subsection{Discrete Floating Strip Micromegas}\label{ssec:discetefsm}
Small Micromegas with an active area of 6.4\,cm\,$\times$\,6.4\,cm have been built with exchangeable 15\,pF SMD capacitors and SMD resistors. This enabled the optimization of recharge resistance with respect to global voltage drop and recharge time. The readout structure consists of 128 copper strips with 300\,$\mu$m width and 500\,$\mu$m pitch.
The readout structure is photolitographically etched from 35\,$\mu$m copper on a standard 1.4\,mm thick FR4 printed circuit board.

The stainless steel micro-mesh is glued onto an aluminum gas frame, that also defines the drift gap. This enables easy cleaning of the amplification gap during assembly. 

Two Micromegas were constructed based on this design: Discharge behavior has been investigated by irradiating a floating strip Micromegas with 6\,mm drift gap with strongly ionizing alpha particles. The results are discussed in detail in \cite{bortfeldt:phd} and demonstrate the excellent discharge behavior of the detector. 
A dedicated floating strip Micromegas with 15\,mm drift region has been constructed for cosmic muon tracking under lateral high-rate proton irradiation.

\subsection{Floating Strip Micromegas with Integrated Capacitors and Resistors}
In a larger prototype with an active area of 48\,cm $\times$ 50\,cm, the anode copper strips are connected to high-voltage using individual paste resistors, signals are extracted by capacitive coupling of the anode strips to a layer of congruent copper readout strips below them. A higher strip density as compared to the discrete type can be achieved. Furthermore the embedded coupling capacitors are intrinsically high-voltage resistant.
\begin{figure}
\centering
\includegraphics[width=0.8\linewidth]{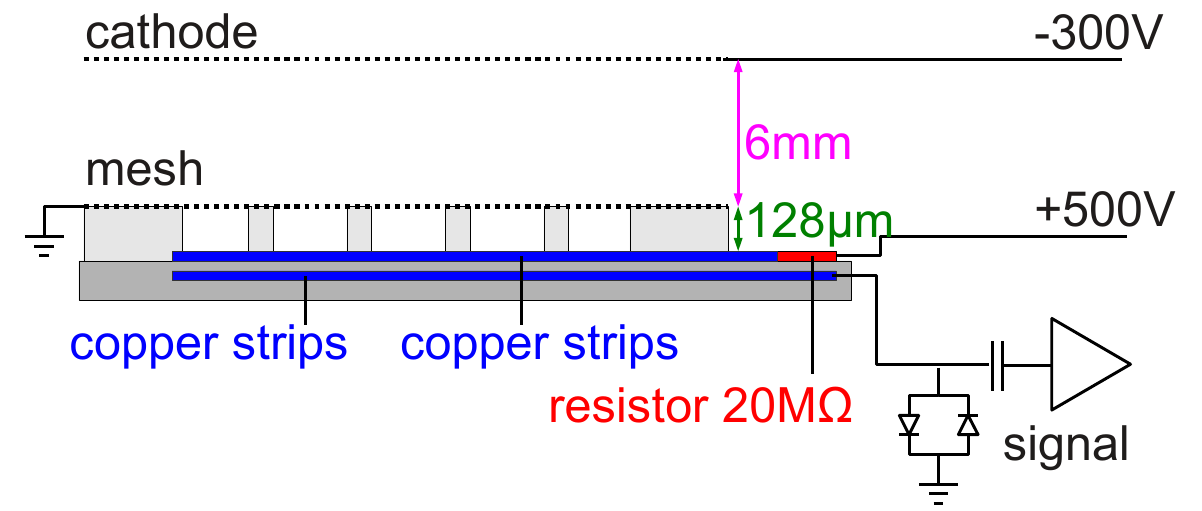}
\caption{Integrated floating strip Micromegas realization.}
\label{fig:FSMicom_Integrated}
\end{figure}
The inner structure of the integrated floating strip Micromegas is shown schematically in Figure\,\ref{fig:FSMicom_Integrated}. The readout structure in this bulk Micromegas consists of 1920 copper strips with 150\,$\mu$m width and a pitch of 250\,$\mu$m. The gas volume is closed by a 1\,mm FR4 panel, carrying the photolithographically formed cathode on its lower side. Since the detector is operated at a relative overpressure of around 30\,mbar, the readout and cathode structure are reinforced by 10\,mm aluminum plates into which 120\,mm $\times$ 120\,mm windows have been cut. Nevertheless a bulging on the order of 2\,mm in the center of the cutouts is visible. It is foreseen to replace the aluminum plates by low-material-budget honeycomb-FR4-panels.

\subsection{Semi-Integrated, Thin Floating Strip Micromegas}\label{ssec:semiintegratedfsm}
\begin{figure}
\centering
\includegraphics[width=0.79\linewidth]{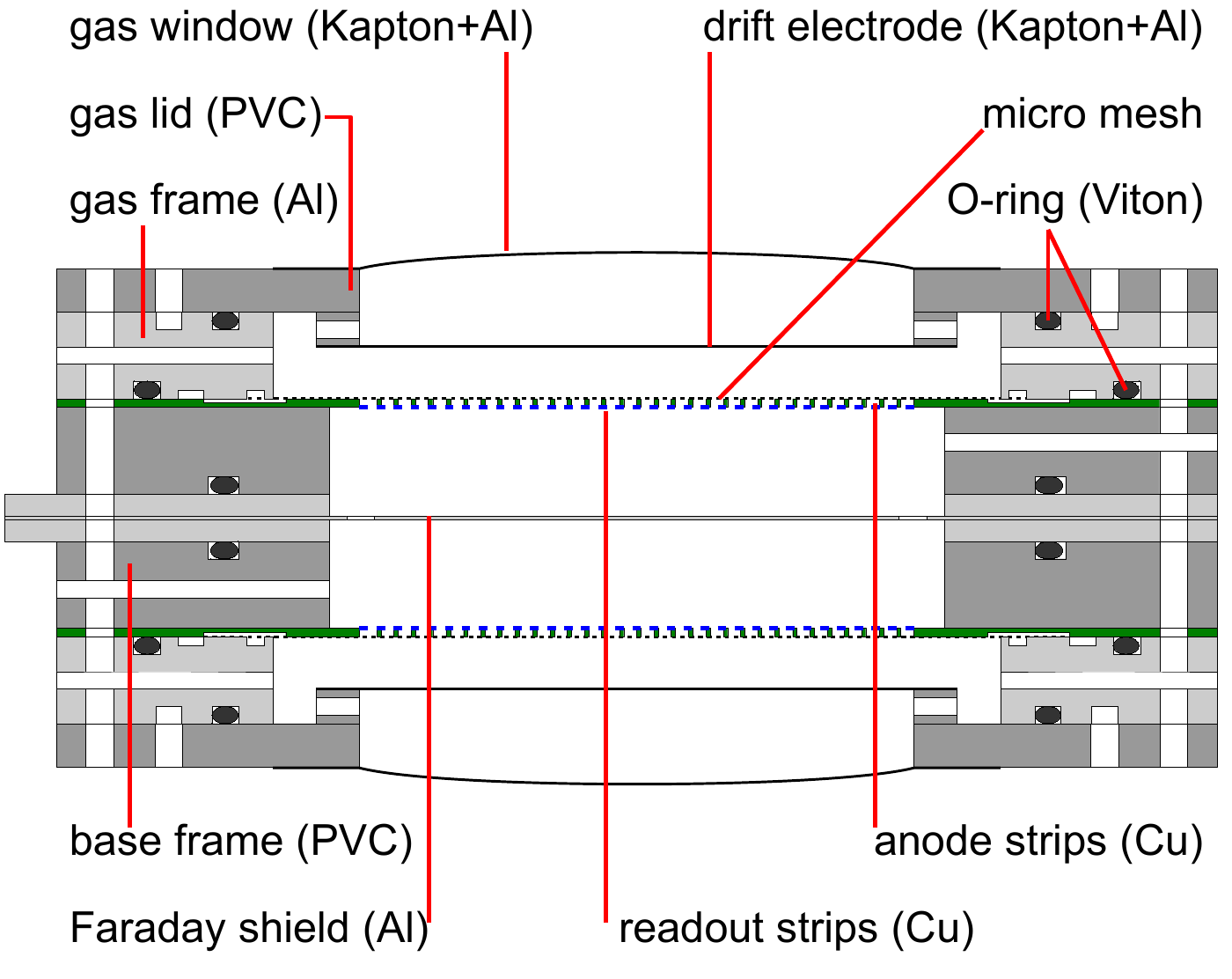}
\caption{Floating strip Micromegas doublet with low material budget.}
\label{fig:FSM_double_Setup_Komment}
\end{figure}

For low-energy ion tracking a 6.4\,cm\,$\times$\,6.4\,cm floating strip Micromegas doublet with very low material budget has been developed. It consists of two Micromegas mounted back-to-back in a single unit. A schematic cut through the detector is shown in Figure \ref{fig:FSM_double_Setup_Komment}.

The gas window as well as the drift electrode consist of aluminized 10\,$\mu$m thick Kapton foils. The woven stainless steel micro-mesh with 400 lines per inch, consisting of 25\,$\mu$m thick wires, is glued to the aluminum gas frame. 128 35\,$\mu$m thick copper anode strips, supported by a 125\,$\mu$m thick FR4 sheet, are individually connected to high-voltage via SMD resistors. Copper readout strips on the lower side of the FR4 material decouple the signal from the anode and are connected to the readout electronics. Anode as well as readout strips exhibit a width of 300\,$\mu$m and a pitch of 500\,$\mu$m. A 30\,$\mu$m thick aluminum foil shields the two sub-detectors from each other. The gas pressure between the two readout structures is dynamically elevated by approximately 8\,mbar with respect to the pressure in the active volume to press the flexible readout structure against the mesh and ensure a homogeneous amplification gap width of around 150\,$\mu$m.

\section{Performance in High-Energy Pion Beams}
In the following we present the performance of the 48\,cm\,$\times$\,50\,cm floating strip Micromegas in 10\,GeV to 160\,GeV pion beams at the H6 beam line at SPS/CERN.

\subsection{Setup}\label{ssec:h6setup}
A beam telescope, consisting of six standard 9\,cm\,$\times$\,10\,cm Micromegas with one dimensional strip readout and two resistive strip 9\,cm\,$\times$\,9\,cm Micromegas with two dimensional strip readout has been used to predict the pion track in the large floating strip Micromegas. The performance of the track telescope and its readout electronics, tested before in muon and pion beams, has been described in \cite{bortfeldt:highrestelescope}. A track accuracy of $\sigma_{\mathrm{track}} = (19.0\pm0.2)\,\mu$m  at the position of the detector under test is reached. Non-resistive detectors were read out with a fast Gassiplex based electronics with 2300 channels, originally developed for the HADES RICH, that has been adapted to Micromegas-like signals. For the resistive and the floating strip detectors APV25 based front-end boards have been used, read out with the RD51 Scalable Readout System \cite{martoiu:srs}. An event rate of 800\,Hz has been reached, limited by the network bandwidth.
 
Offline synchronization of the two readout systems and jitter reduction is enabled by acquiring a binary 12 bit trigger number created in a custom event counter, with an additional Time-to-Digital converter in the Gassiplex and unused input channels in the APV25 system. 

\subsection{Single Plane Track Inclination Reconstruction}\label{ssec:utpc}
We consider a pion, traversing the detector at a non-zero angle with respect to the direction of the drift field lines. By measuring the arrival time of ionization electrons on the readout strips, their drift distance and thus their creation point can be calculated using field dependent electron drift velocities, determined with MAGBOLTZ \cite{biagi:magboltz}. In combination with the strip position, this yields two-dimensional space points, that can be fitted with a straight line to directly determine the track inclination angle. Due to the analogy with Time-Projection-Chambers, this method is called $\mu$TPC reconstruction.

\begin{figure}
\centering
\includegraphics[width=0.7\linewidth]{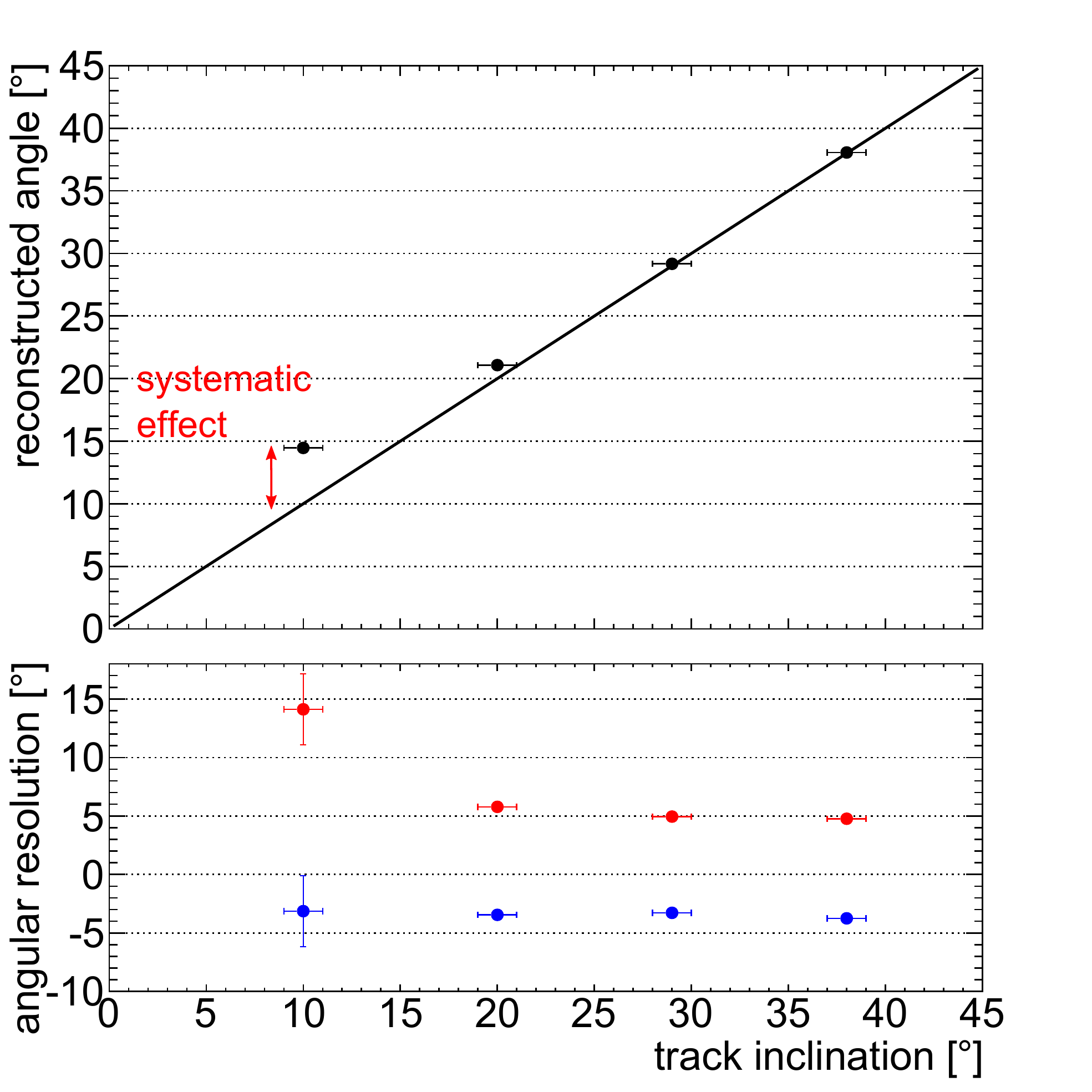}
\caption{Reconstructed track inclination (top) and angular resolution (bottom) as a function of the true track inclination, measured with  $E_\mathrm{amp}=37.5\,$kV/cm and $E_\mathrm{drift}=0.3\,$kV/cm.}
\label{fig:recanglevsangle}
\end{figure}

In Figure \ref{fig:recanglevsangle}, the most probable reconstructed track inclination and the angular resolution are shown as a function of the true track inclination.
The reconstructed track inclination angle is systematically shifted towards larger values with respect to the true track inclination by two effects: First, signals couple capacitively onto neighboring strips. Thus measured signal timing is shifted towards earlier values for the strips, hit by electrons, created close to the cathode. Second, for strips at the edges of the strip cluster, the ionization electrons position is not calculated correctly. The term cluster denotes in this paper a group of adjacent strips, registering the signal of a certain particle.

The systematic deviations decrease with increasing track inclination. They have been studied in detail and are quantitatively understood, details can be found in \cite{bortfeldt:phd}.

\subsection{Spatial Resolution, Pulse Height and Efficiency}
For perpendicular incidence, the measured hit position can be reliably calculated from the charge-weighted mean of the strip positions in the cluster. We determine the difference between the measured particle position and the position predicted by the reference detectors for many congeneric tracks and extract the width $\sigma_\mathrm{resid}$ of this residual distribution by fitting with a double Gaussian function. The spatial resolution is then calculated by quadratically subtracting the track accuracy.
\begin{figure}
\centering
\includegraphics[width=0.7\linewidth]{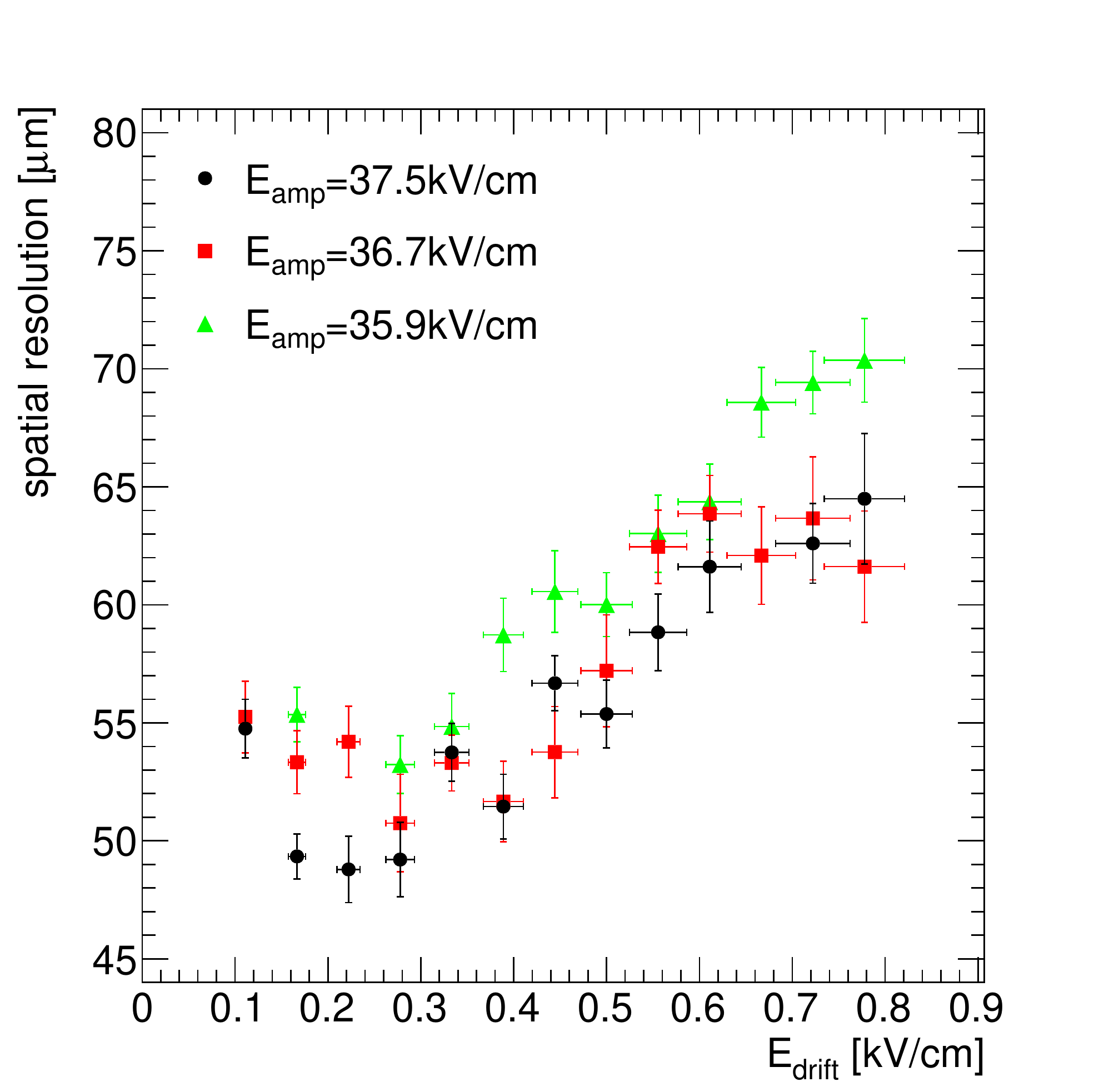}
\caption{Spatial resolution as a function of the drift field for different amplification field, measured with perpendicularly incident 120\,GeV pions.}
\label{fig:spatresvsed}
\end{figure}
The spatial resolution as a function of the drift field is shown in Figure \ref{fig:spatresvsed} for different amplification fields. It reaches optimum values at $0.2\,\mathrm{kV/cm}\lesssim E_\mathrm{drift}\lesssim 0.3\,$kV/cm, where the transverse electron diffusion is minimal. Note, that the dependance of the spatial resolution on the transverse electron diffusion is larger than its dependance on the absolute pulse height.

The pulse height shows the expected dependance on the drift field, c.f.~\cite{bortfeldt:highrestelescope}: Starting at small fields, the pulse height increases due to an improved separation of ionization charge and a increasing electron drift velocity. Both effects lead to an increase of the charge, detectable during the preamplifier shaping time. Due to an increasing mesh opacity, caused by reconfiguration of the electric field lines and the increasing transverse electron diffusion, the pulse height decreases for $E_\mathrm{drift}>0.5\,$kV/cm. 

The hit efficiency is calculated by selecting events in which all reference detectors registered a hit and then testing whether the detector under test has also been hit. The measured efficiency of the floating strip Micromegas increases with increasing drift field and reaches a plateau for $E_\mathrm{drift}\gtrsim0.25\,$kV/cm. Optimum values above 0.95 are reached for the highest amplification field.

Pulse height, efficiency and spatial resolution have been measured at three different positions in the floating strip Micromegas: In the center and close to the two upper edges. The measured efficiency and spatial resolution show no position dependance, the measured pulse heights show spatial variations below 20\%.

\section{Performance in a High-Rate Background Environment}
The cosmic muon tracking capabilities of a discrete floating strip Micromegas under lateral irradiation with 20\,MeV protons at a rate of 550\,kHz at the Tandem accelerator in Garching is discussed in the following.

\subsection{Setup}
The floating strip Micromegas under test has been described in section \ref{ssec:discetefsm}. Two opposing 6.4\,cm\,$\times$\,0.9\,cm cutouts in the aluminum frame, closed with 10\,$\mu$m Kapton foil, allowed for lateral irradiation with 20\,MeV protons. To reduce the material budget, the 0.5\,mm FR4 readout board is supported by a sandwich of 10\,mm thick aluminum honeycomb and bare FR4 material. Flatness is ensured by manufacturing the supportive structure on a precise granite table.

\begin{figure}
\centering
\includegraphics[width=0.87\linewidth]{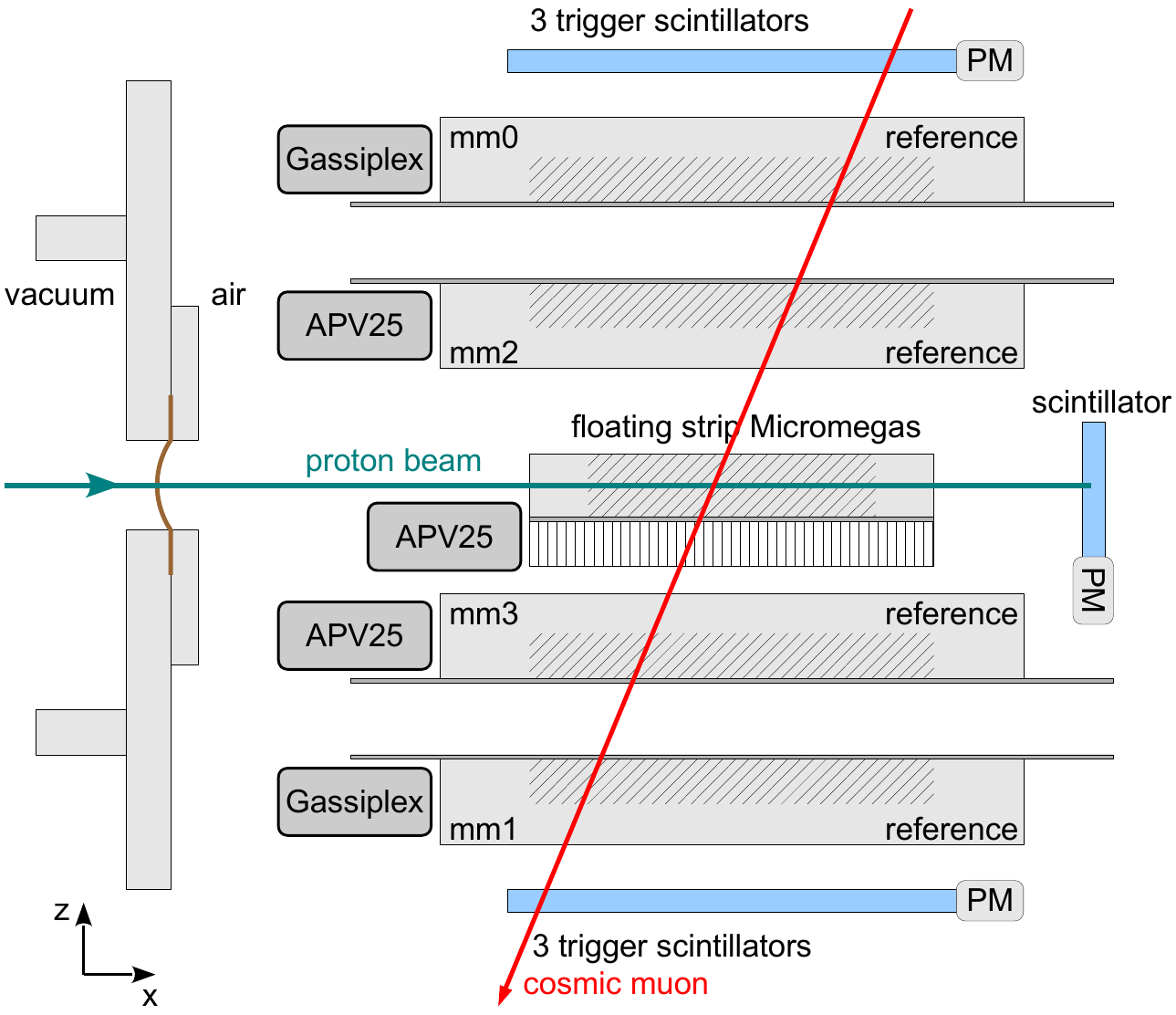}
\caption{Cosmic muon telescope at the Tandem accelerator in Garching.}
\label{fig:tandem_mtrack_setup}
\end{figure}

Four Micromegas are used to reference the cosmic muon track parameters, two standard Micromegas with one-dimensional strip readout and two resistive strip Micromegas with two-dimensional strip readout, Figure \ref{fig:tandem_mtrack_setup}. The reference detectors feature an active area of 10\,cm\,$\times$\,9\,cm and 9\,cm\,$\times$\,9\,cm, respectively, and a strip pitch of 250\,$\mu$m, enabling a mean reference track accuracy of $\sigma_{\mathrm{track}} = (50.4\pm0.5)\,\mu$m. 

Gassiplex and APV25 based system are used to read out the Micromegas, c.f.~section \ref{ssec:h6setup}. Coincident muon hits in two scintillator layers trigger the data acquisition system.

Traversing protons can produce a signal on all readout strips in the irradiated floating strip Micromegas, thus representing the worst possible background situation. They are stopped behind the Micromegas in a scintillator detector that allows for accurately monitoring the actual proton rate. Using the number of hit clusters in the floating strip Micromegas as discriminator, events in which a proton crossed the detector in coincidence with the triggering muon can be identified. The measured ratio of these events is $P_{\mu + p}=(0.39\pm0.02)$ and agrees well with the expectation.

\subsection{Spatial Resolution, Efficiency and Stability}
\begin{figure}
\centering
\includegraphics[width=0.7\linewidth]{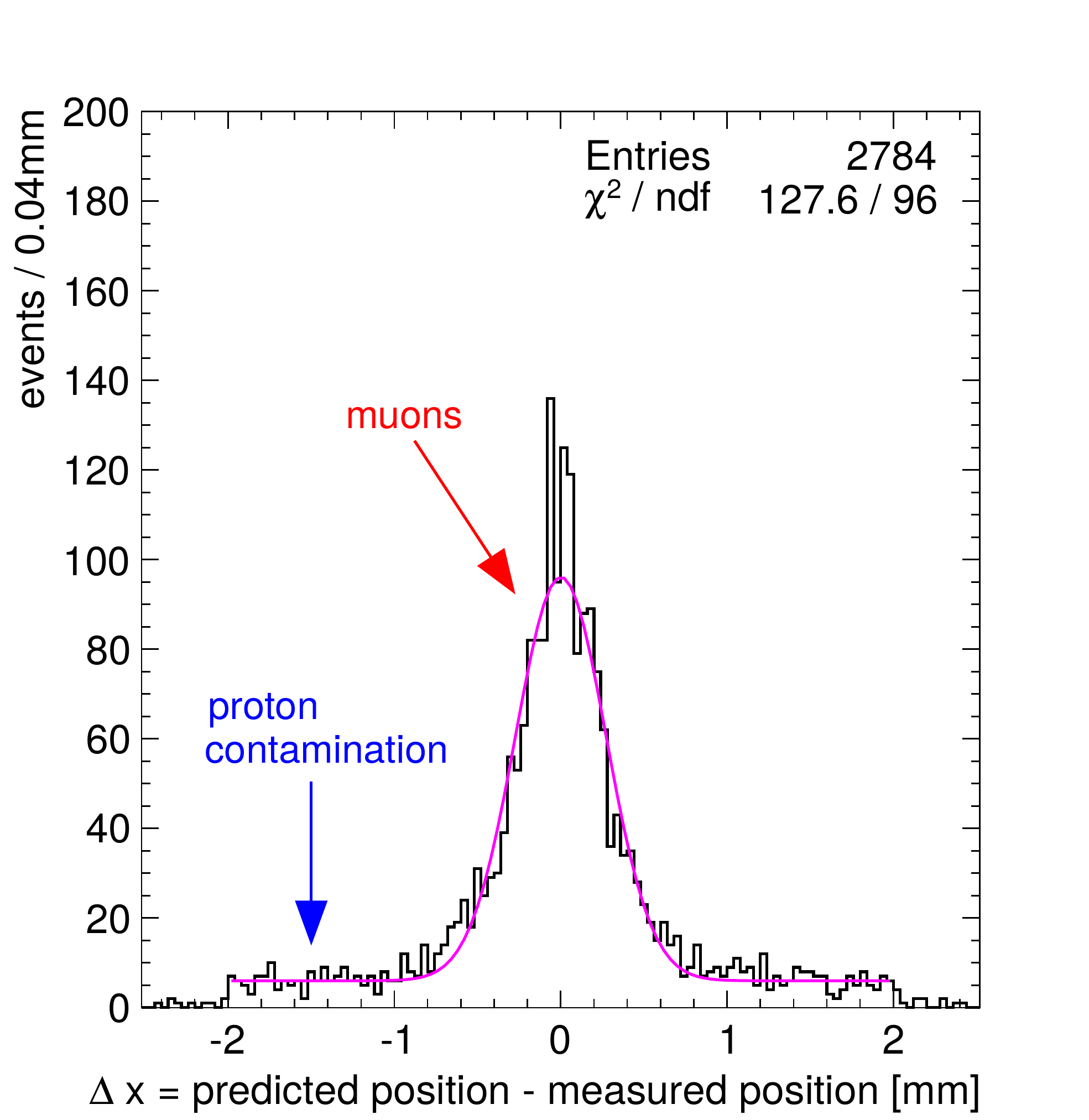}
\caption{Residual between measured and predicted cosmic muon hit position in the floating strip Micromegas under proton irradiation.}
\label{fig:Res4Fit_9998_mtrack}
\end{figure}

\begin{figure}
\centering
\includegraphics[width=0.7\linewidth]{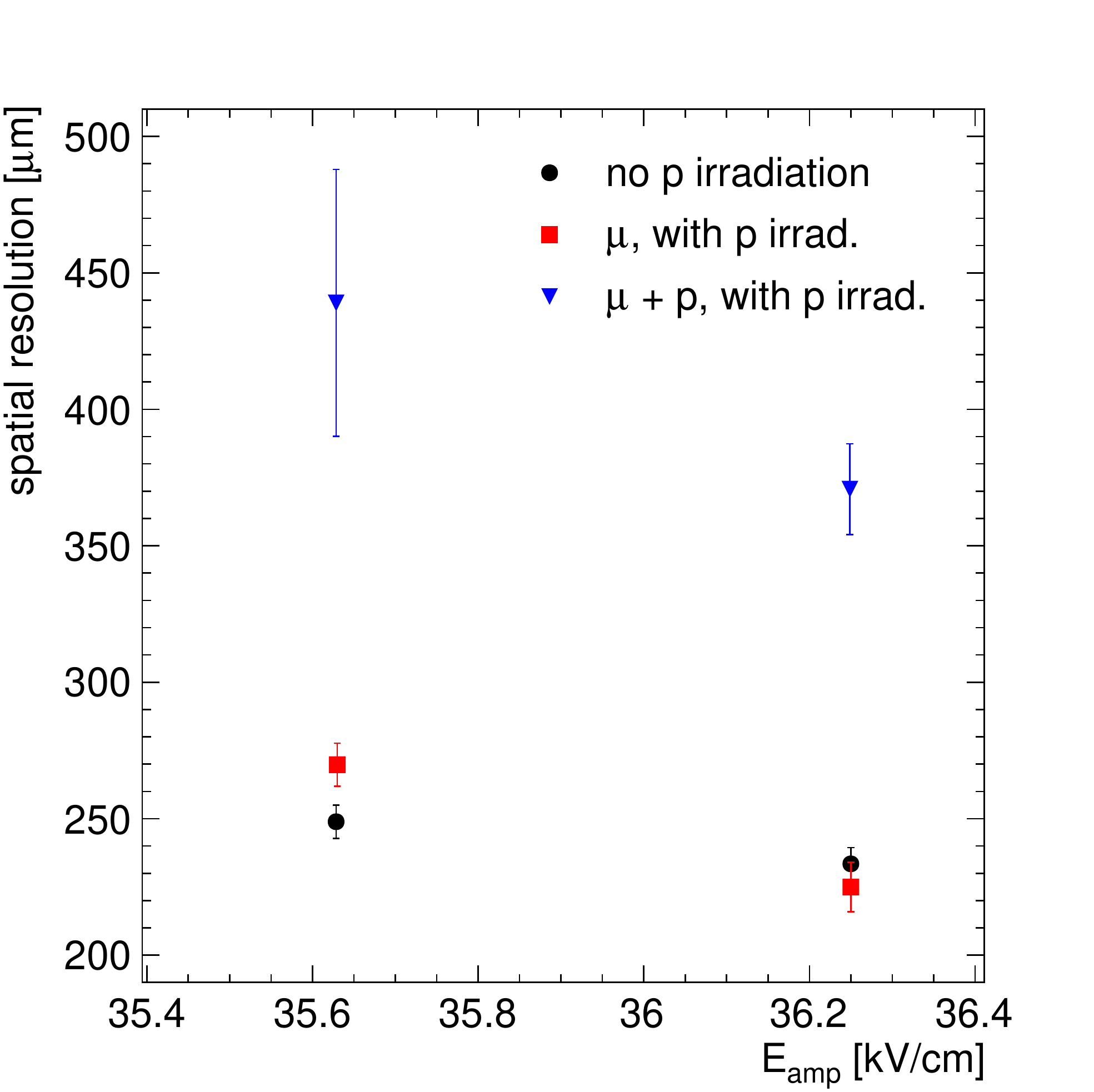}
\caption{Spatial resolution at two different amplification fields. Black circles correspond to a measurement without proton irradiation. The measurement with irradiation is further separated into events with only a cosmic muon (red squares) and with a cosmic muon and a coincident proton in the respective trigger window (blue triangles).}
\label{fig:SpatRes4VsEamp4_tanAug13_mtrack}
\end{figure}

The cluster, closest to the predicted muon position in the floating strip Micromegas, is attributed to the cosmic muon. The floating strip Micromegas is not included in the track reconstruction algorithm to prevent biasing. If no cluster is found in a $\pm2\,$mm region around the predicted position, the detector is considered as inefficient. Figure \ref{fig:Res4Fit_9998_mtrack} shows the distribution of residuals between measured and predicted muon position in the floating strip Micromegas. Correctly reconstructed cosmic muons form a narrow peak on a flat background, caused by proton and noise signals. 
The distribution is fitted with the sum of a Gaussian function and a constant. The Gaussian standard deviation allows for calculating the spatial resolution of the floating strip Micromegas (Figure \ref{fig:SpatRes4VsEamp4_tanAug13_mtrack}), the proton contamination can be estimated from the constant. 

The spatial resolution under irradiation in events with only a cosmic muon would be a sensitive measure for indirect effects of irradiation such as field distortion or space charge production. Indirect effects can be excluded in the presented measurement, as the spatial resolution agrees well with the corresponding value without irradiation. Only in events with coincident muon and proton signals in the same strip region, the muon reconstruction is distorted.

If proton signals would completely conceal coincident muon signals, a ratio $\varepsilon_{\mathrm{irrad}}/\varepsilon_{\mathrm{no\,irrad}} = (0.62\pm0.01)$, between the muon detection efficiency in measurements without and with additional proton irradiation would be expected. 
The observed ratio of $\varepsilon_{\mathrm{irrad}}/\varepsilon_{\mathrm{no\,irrad}} = (0.71\pm0.01)$ indicates, that muon identification is to some extent possible even with coincident proton signals.

The discharge rate of the  floating strip Micromegas has been monitored by registering the small recharge current after a discharge. A mean discharge rate of $(0.173\pm0.002)\,$Hz leads to a negligible inefficiency of $4.1\times10^{-6}$.

\section{High-Rate Ion Tracking with Thin Floating Strip Micromegas}
The performance of a low-material-budget floating strip Micromegas doublet has been investigated at the Heidelberg Ion Therapy center.

\subsection{Setup}
\begin{figure}
\centering
\includegraphics[width=0.82\linewidth]{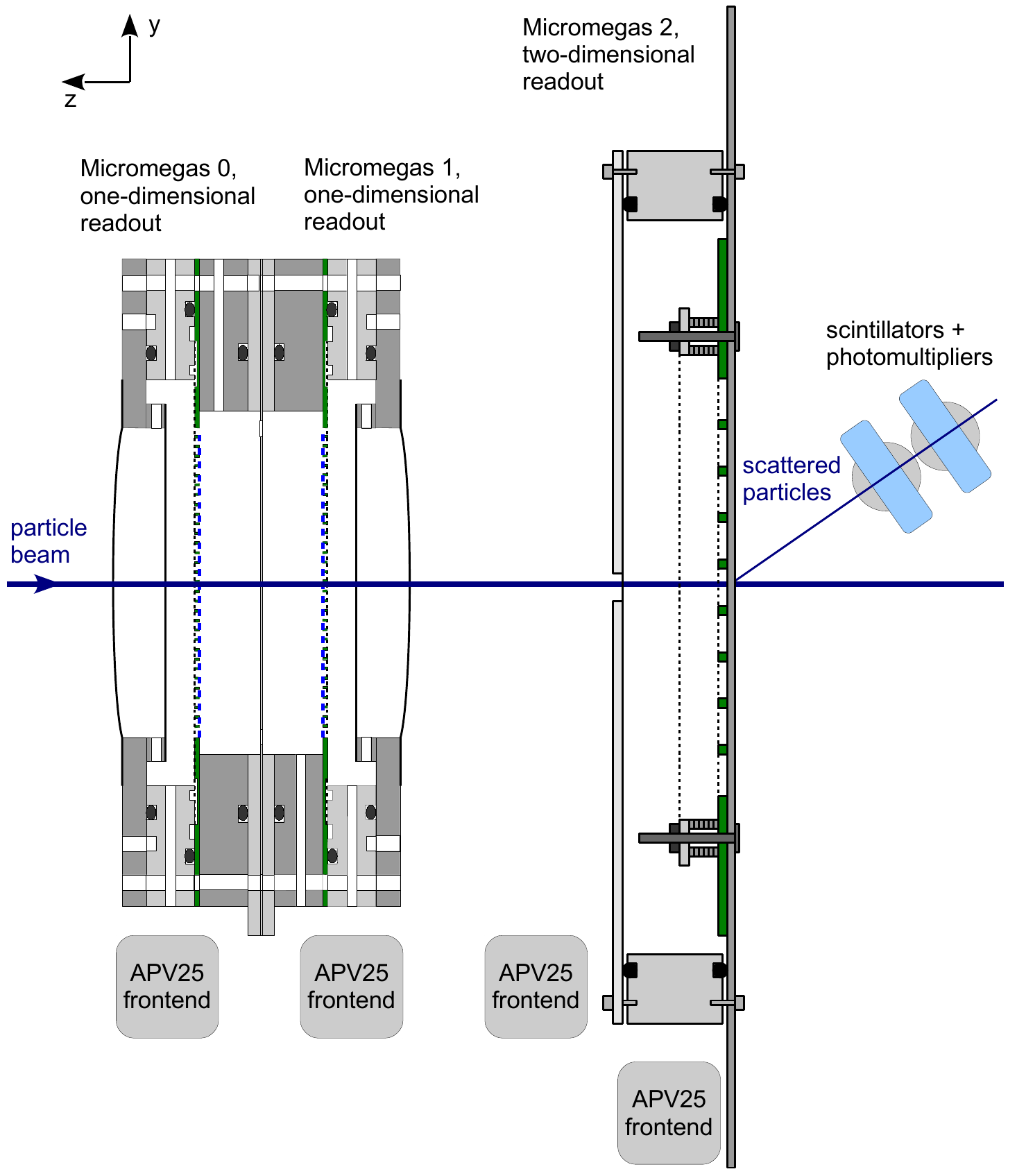}
\caption{Setup with three Micromegas at the HIT, the trigger was provided by a set of scintillator detectors, sensitive to scattered charged particles.}
\label{fig:hit_setup}
\end{figure}

The setup is shown schematically in Figure \ref{fig:hit_setup}. The floating strip Micromegas doublet, described in section \ref{ssec:semiintegratedfsm}, and a resistive strip Micromegas with two-dimensional strip structure were read out with APV25 based electronics.
$^{12}C$ beams with $E_{\mathrm{ion}}= 88\,$MeV/u to 430\,MeV/u were available at particle rates between 2\,MHz and 80\,MHz as well as proton beams with  $E_{\mathrm{p}}= 48\,$MeV to 221\,MeV at rates between 80\,MHz and 2\,GHz. 

\subsection{Rate Behavior of the Pulse Height, Spatial Resolution and Efficiency}
At particle rates around 10\,MHz, corresponding to flux densities of 7\,MHz/cm$^2$, the reconstructed number of particle hits per event is on the order of 4 and agrees with the expected number of true hits. For further increasing particle rate the mean number of reconstructed hits saturates at around 5.5 as hits by different particles merge into conjoint clusters.

\begin{figure}
\centering
\includegraphics[width=0.7\linewidth]{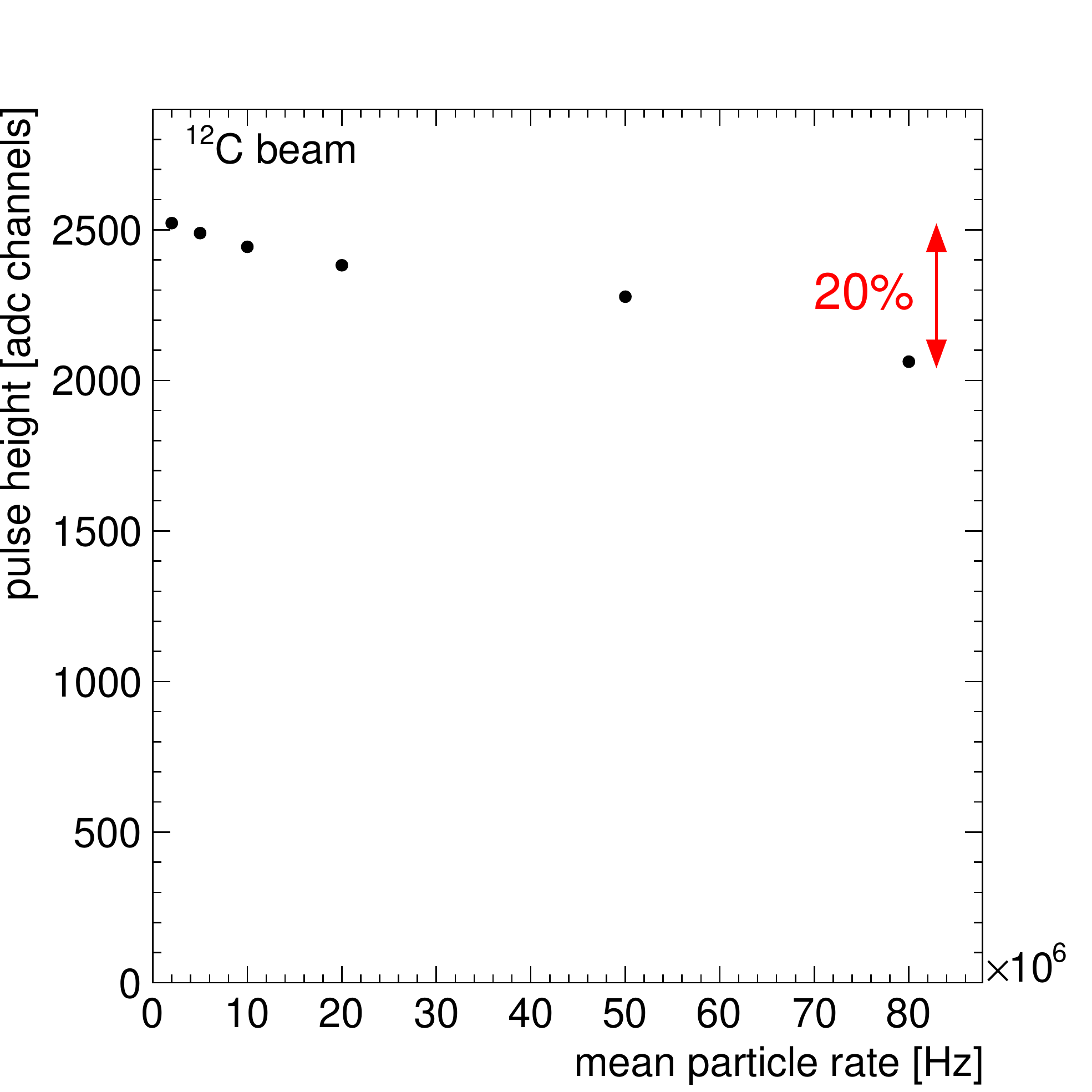}
\caption{Pulse height of the first floating strip Micromegas for carbon ions at $E_{\mathrm{ion}}= 88.83\,$MeV/u. Only a drop of 5\% could in principle not be compensated for.}
\label{fig:Ph0_1_2VsRate_C12}
\end{figure}

Single particle hits can still be identified at the highest carbon ion rates, corresponding to flux densities of 60\,MHz/cm$^2$, although not all particle tracks can be separated. The single particle pulse height as a function of the rate is shown in Figure \ref{fig:Ph0_1_2VsRate_C12}.
The pulse height decreases by 20\% when increasing the rate from 2\,MHz to 80\,MHz. A 15\% drop is explained by the 4\,V voltage drop over the individual recharge resistors of the floating strip Micromegas, that is caused by the measured recharge current of 180\,nA per strip. During the beam time we refrained from the possible compensation of this drop by adjustment of the high-voltage supply. The remaining pulse height decrease on the order of 5\% is attributed to space charge effects in the amplification region.

\begin{figure}
\centering
\includegraphics[width=0.7\linewidth]{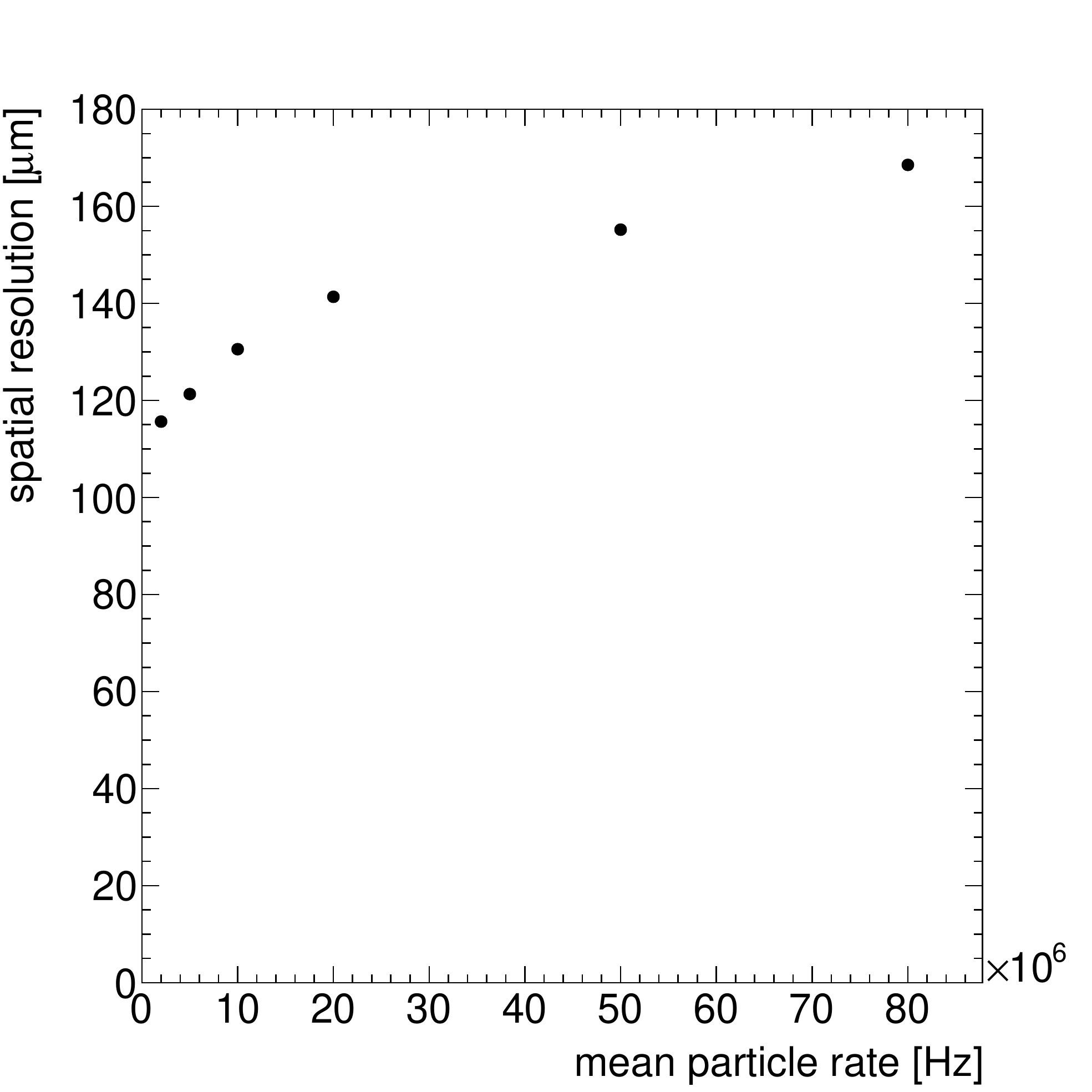}
\caption{Spatial resolution of the first floating strip Micromegas for carbon ions at $E_{\mathrm{ion}}= 88.83\,$MeV/u. It is determined with a geometric mean method, discussed in \cite{Carnegie:SpatRes} and \cite{bortfeldt:highrestelescope}.}
\label{fig:SpatRes0VsRate_C12_hitcorrect}
\end{figure}

The rate dependance of the spatial resolution of the first floating strip Micromegas is shown in Figure \ref{fig:SpatRes0VsRate_C12_hitcorrect}.
The visible degradation of the spatial resolution with particle rate is caused by deterioration of the reconstructed particle position by coincident hits on the same or adjacent strips. Even at the highest rates achievable with carbon ions though, a spatial resolution well below 200\,$\mu$m can be achieved, completely sufficient for the currently intended application in medical ion transmission radio\-graphy and tomography.

The detection efficiency of the first floating strip Micromegas for carbon ions with  $E_{\mathrm{ion}}= 88.83\,$MeV/u is for all particle rates above 0.99 and for protons with  $E_{\mathrm{p}}= 221.06\,$MeV for all particle rates above 0.97. The detector overall availability is above 0.97 even at the highest rates of 2\,GHz despite of corresponding discharge rates around 40\,Hz. 

\section{Conclusion}
Optimized, discharge tolerant floating strip Micromegas were presented. The performance of 6.4\,cm\,$\times$\,6.4\,cm and 48\,cm\,$\times$\,50\,cm floating strip detectors in high-energy pion beams, for cosmic muon tracking under high-rate background irradiation and in highest-rate ion beams has been discussed. We have demonstrated the  good discharge insensitivity and have shown that the excellent performance of the detectors with respect to spatial resolution and detection efficiency is preserved up to the highest particle rates. This substantiates that floating strip Micromegas are versatile particle tracking detectors, suited for various applications such as muon tracking in high energy physics and low-energy ion tracking in medical physics.

\section*{Acknowledgments}
The authors thank M.~B{\"o}hmer and L.~Maier, TU Munich, for providing the Gassiplex readout and S.~Brons and the HIT accelerator team for the support.
This work was supported by the DFG cluster of excellence on ``Origin and Structure of the Universe'' and the DFG research training group 1054.




\bibliographystyle{elsarticle-num}

\end{document}